\documentstyle[12pt,epsf]{article}
\def\simle{\hspace*{0.2em}\raisebox{0.5ex}{$<$}
     \hspace{-0.8em}\raisebox{-0.3em}{$\sim$}\hspace*{0.2em}}
\newcommand{\beq}{\begin{equation}}
\newcommand{\eeq}{\end{equation}}
\newcommand{\beqa}{\begin{eqnarray}}
\newcommand{\eeqa}{\end{eqnarray}}

\newcommand{\R}{{\cal R}}
\newcommand{\Rp}{{\cal R'}}

\begin{document}

\begin{titlepage}

\hfill{NT@UW-00-023}

\hfill{DOE/ER/411332-102-INT00}

\hfill{RBRC-140}

\hfill{KRL MAP-271}

\vspace{1.0cm}

\begin{center}
{\Large\bf Singular Potentials and Limit Cycles}

\vspace{1.2cm}

{\bf S.R.~Beane}$^{a,}$\footnote{{\tt sbeane@phys.washington.edu}},
{\bf P.F.~Bedaque}$^{b,}$\footnote{{\tt bedaque@phys.washington.edu}},
{\bf L.~Childress}$^{a,}$\footnote{{\tt childres@fas.harvard.edu}},\\
{\bf A.~Kryjevski}$^{b,}$\footnote{{\tt abk4@u.washington.edu}},
{\bf J.~McGuire}$^{b,}$\footnote{{\tt james.mcguire@yale.edu}}, and
{\bf U.~van Kolck}$^{c,d,e,}$\footnote{{\tt vankolck@krl.caltech.edu}}

\vspace{0.5cm}
{\it
$^a$Department of Physics and $^b$Institute for Nuclear Theory,\\
 University of Washington,\\
 Seattle, WA 98195\\
~\\$^c$ Department of Physics,\\
 University of Arizona,\\
 Tucson, AZ  85721\\
~\\$^d$ RIKEN-BNL Research Center,\\
 Brookhaven National Laboratory,\\
Upton, NY 11973\\
~\\$^e$ Kellogg Radiation Laboratory, 106-38,\\
 California Institute of Technology, \\
 Pasadena, CA 91125}
\end{center}

\vspace{0.5cm}

\begin{abstract}
We show that a central $1/r^n$ singular potential (with $n\geq 2$) is
renormalized by a one-parameter square-well counterterm; low-energy
observables are made independent of the square-well width by adjusting
the square-well strength.  We find a closed form expression for the
renormalization-group evolution of the square-well counterterm.
\end{abstract}

\vspace{2cm}
\vfill
\end{titlepage}

\setcounter{page}{1}

\section{Introduction}

The study of singular potentials in quantum mechanics is almost as old
as quantum mechanics itself~\cite{plesset}. Physically, singular
potentials pose problems because the force between two particles,
represented by the potential, does not uniquely determine the
scattering problem~\cite{case}. Here we will focus on singular
potentials of $1/r^n$ type, with $n\geq 2$.  Classically, particles
subject to such a force fall to the origin with an infinite
velocity. In the quantum theory, the wavefunction oscillates
indefinitely on the way to the origin, allowing no way of specifying a
linear combination of solutions~\cite{case}.  Of course, in any
physical situation described by a singular potential, the potential is
intended as a description of long-range behavior, so there is a sense
in which the pathologies which occur near the origin are irrelevant to
the physical problem. This should remind the reader of the infinities
encountered in quantum field theory which are cured through
renormalization. This analogy with field theory has provided an
important motivation for the study of singular
potentials~\cite{singpotreview}.

If a singular potential itself is not sufficient to determine the
scattering problem, one might be tempted to classify the singular
potentials as nonrenormalizable and abandon all hope.  This point of
view is now outdated. In the modern version of the renormalization
paradigm a low-energy system with a clear-cut separation of scales can
be described by an effective field theory (EFT) involving explicitly
only the long-wavelength degrees of freedom, and organized as an
expansion in powers of momenta~\cite{dave}. The short-range dynamics
can always be treated as a set of local operators.  In the present
context, the $1/r^n$ potential represents the long-distance part of
the potential. Local operators in momentum space correspond to
delta-function interactions in coordinate space. The essential point
of EFT is that the details of the short-distance physics are not of
importance to low-energy scattering. Hence one can simulate the delta
function in an infinite number of ways. The simplest choice of a
``smeared out'' delta function is a simple square well. With a
singular potential representing a given long-distance force, and a
square well representing unknown short-distance physics, an
interesting question is whether one can obtain an EFT with well
defined low-energy scattering observables, which are to a specified
degree of accuracy insensitive to the short-distance physics encoded
by the square well. It is the purpose of this paper to explore this
issue. Note that we do not attempt to renormalize the coupling
strength of the singular potential itself~\cite{camblong}. In the
physical problems of interest, the coupling strength is completely
determined by the long-distance physics so there is no freedom to
renormalize this parameter.

By way of physical motivation we note that the singular potentials of
$1/r^n$ type are of great current physical interest. The special case
$n=2$ is relevant to the three-body problem in nuclear
physics~\cite{efimov}\cite{bosons}\cite{wilson}. This case is also 
relevant to point-dipole interactions in molecular
physics~\cite{leblond}. The case $n=3$ corresponds to the tensor force
between nucleons and is at the heart of nuclear physics. The issue of
the proper renormalization of this potential is an essential
ingredient of the intense ongoing effort to develop a perturbative
theory of nuclear interactions~\cite{monster}. The interaction between
a charge and an induced dipole is of type $n=4$~\cite{vogt}. The case
$n=5$ is a perturbative correction to the tensor force in the nuclear
potential~\cite{monster}. Both $n=6$ and $n=7$ correspond to van der
Waals forces, of London~\cite{landau} and Casimir-Polder~\cite{vander}
type, respectively.

This paper is organized as follows. In section 2 we set up the quantum
mechanical scattering problem of two particles subject to a $1/r^n$ potential
with a square well.  In section 3 we consider the marginal $n=2$ case in some
detail. The pure singular potentials $n\geq 3$ are considered at zero energy in
section 4. In section 5 we make use of the WKB approximation to generalize our
results to non-zero energy and to estimate the errors associated with the
renormalization procedure.  We discuss the applicability of a perturbative
expansion for singular potentials in section 6. Our numerical analysis, for the
case $n=4$, is discussed in section 7. We discuss and conclude in section 8.

\section{The $1/{r^n}$ potential with a Square Well}

We consider two particles of reduced mass $M$ interacting in the $S$ wave with
a singular potential that goes as $1/r^n$, $n\geq 2$.  This potential has a
scale that sets its curvature, $r_0$; this is the characteristic scale of the
long-distance physics.  The strength of the long-distance potential is governed
by a parameter $\lambda_L/2M r_0^2$.  To obtain well-defined solutions, we need
to regulate the potential by introducing a cutoff procedure.  Since we are
posing the problem in coordinate space, we do this through a cutoff radius
$R\equiv{\cal R} r_0$, ${\cal R}\simle 1$.  We expect that the solutions will
depend sensitively on ${\cal R}$, that is, on the short-range physics.  We
simulate a short-range delta-function interaction by a square well of this
radius and with depth $\lambda_S/2M r_0^2$.  The problem will be correctly
renormalized once we are able to vary ${\cal R}$ (inasmuch as ${\cal R}\simle
1$) and simultaneously $\lambda_S$ in such a way as to keep observables (say
phase shifts) invariant. The corresponding constraint
$\lambda_S=\lambda_S({\cal R})$ represents the renormalization-group flow of
the contact interaction~\footnote{Here we mean that there is a ``group'' of
  transformations on the cutoff $\cal R$ which leaves observables invariant.}.
We will see that this short-distance physics is represented by the running
coupling

\begin{equation}
H_n ({\cal R})\equiv\sqrt{\lambda_S({\cal R})}{\cal R}.
\end{equation}

We thus take as our potential

\begin{equation}
V(r)=\frac{1}{2Mr_0^2}\biggl(-\lambda_S\theta (R-r) - \lambda_L\frac{f(r/r_0)}{(r/r_0)^n} \theta (r-R)
\biggr),
\end{equation}
where $f(x)$ is a regular function of $x$ near the origin with $f(0)=1$,
$f(1)={\cal O}(1)$. Notice that $\lambda_S,\lambda_L>0$ correspond to purely
attractive potentials.  In terms of $x=r/r_0$, the Schr\"odinger equation for
the wavefunction $u(r)/r$ at an energy $E=k^2/2M =\eta^2/2M r_0^2$ is

\begin{equation}
\left\{ \begin{array}{ll}
        u''(x)+ (\eta^2+\lambda_S)u(x)=0 & x < {\cal R} \\
        u''(x)+ (\eta^2+\lambda_L\frac{f(x)}{x^n})u(x)=0 
                                                        & x > {\cal R}.
        \end{array} \right. \label{sch1}
\end{equation}
We will consider the simplified case
$f(x)=1$ until section 5.  

There is a very simple argument for classifying singular potentials which we
will repeat here~\cite{landau}. In the vicinity of the origin the uncertainty
principle dictates that the kinetic energy scales like ${x^{-2}}$. Therefore in
a system described by an attractive singular potential alone, the Hamiltonian
of the system is given by the sum of the kinetic energy and the potential
energy, $-{\lambda_L}{x^{-n}}$.  Note that for a Coulomb potential, $n=1$, and
for a sufficiently weak $n=2$ potential, the Hamiltonian is bounded and
therefore the Schr\"odinger equation has a unique regular solution.  Clearly
for $n=2$ and $\lambda_L$ sufficiently strong, the Hamiltonian is unbounded
from below.  Furthermore, when $n\geq 3$ the Hamiltonian is always unbounded
from below.  Hence, an attractive singular potential alone is meaningless in
the vicinity of the origin; the unboundedness of the Hamiltonian represents the
onset of short-distance physics whose effect must be included in the potential.

\section{The Marginal Case: $n=2$}

\subsection{The $k=0$ Solution}

Consider first the zero-energy solution $u(x;0)$. 
The Schr\"odinger equation for
$x>{\cal R}$ is

\begin{equation}
        u''(x;0)+ {{\lambda_L}\over{x^2}}u(x;0)=0 
\end{equation}
and the general solution is

\begin{equation}
        u(x;0)=A x^{{1\over 2}+\gamma}+ B x^{{1\over 2}-\gamma}
\label{marginalzeroen}
\end{equation}
where $\gamma=\sqrt{{1/4-\lambda_L}}$.  For $\lambda_L < 1/4$ the
solution is well known~\cite{landau} and will not be further
considered in this paper. On the other hand, for $\lambda_L > 1/4$ we
can define $\gamma\equiv i\nu$, and the general solution is

\begin{equation}
        u(x;0)= \sqrt{x}\cos{\left( \nu\log{(x)}+\phi_2 \right)}
\label{marginalsmx}
\end{equation}
where $\nu =\sqrt{{\lambda_L-1/4}}$, $\phi_2 =(\log{A/B})/2i$ and we ignore the
overall normalization. Both of the linearly independent solutions of
Eq.~(\ref{marginalzeroen}) vanish as $x\rightarrow 0$, and oscillate
indefinitely on the way there. There is no obvious way to determine a unique
linear combination of solutions; i.e. fix $\phi_2$.  This is the fundamental
problem with singular potentials in quantum mechanics. Renormalization theory
tells us that this sickness is to be expected and arises from probing
arbitrarily short-distance scales, where the true potential no longer has the
form $1/x^2$.  The cure is to cut off the long-distance potential at a radius
${\cal R}$ and introduce a simple parametrization of the unknown short-distance
physics, since low-energy observables cannot distinguish between the
schematic, parametrized potential and the true potential at short distances. We
choose a square well for simplicity, but we emphasize that any choice of
function is equally valid.

\subsection{Matching to the Square Well}

\begin{figure}[t]
\centerline{{\epsfxsize=4in \epsfbox{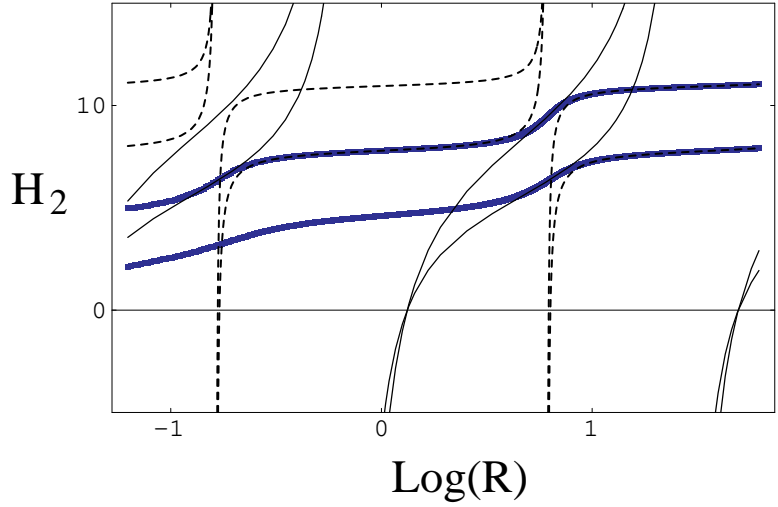}} }
\noindent
\caption{\it 
The running coupling for the $n=2$ singular potential. 
The solid lines are given by Eq.~(\ref{marginalmatchresult}) and
the dashed lines are given by Eq.~(\ref{marginalmatchresult2}). 
The bold lines are a numerical solution of Eq.~(\ref{marginallogder}).}
\label{running2}
\vskip .2in
\end{figure}

The solution in the interior region, $x<{\cal R}$, is straightforward.
It is sufficient to consider an attractive square-well potential,
$\lambda_S >0$.  Matching logarithmic derivatives at the boundary
$x={\cal R}$ gives

\begin{equation}
\sqrt{\lambda_S}\cot{\sqrt{\lambda_S }{\cal R}}={1\over {\cal R}}
\left\{ {1\over 2} - 
{\nu}\tan\left( {{\nu}}\log{({\cal R})}+\phi_2\right)\right\}.
\label{marginallogder}
\end{equation}
If we vary ${\cal R}$ and $\lambda_S({\cal R})$ as given here, the zero-energy
phase $\phi_2$ will not be affected.  Eq.~(\ref{marginallogder}) is
transcendental and therefore rather cumbersome.  However there are two regimes
where an analytical expression can be found. The first is when
$\cot{\sqrt{\lambda_S }{\cal R}}$ is large.  This is the generic situation as
${\cal R}\rightarrow 0$ since the the right-hand side of this equation blows
up, except where $ {\nu}\tan( {{\nu}}\log{({\cal R})}+\phi_2)=0 $.  We can then
find an approximate solution by writing $\sqrt{\lambda_S }{\cal R} =m\pi
+\epsilon$ where $m$ is an integer and $\epsilon$ is a small number. Inserting
this into Eq.~(\ref{marginallogder}) and keeping the leading order in
$\epsilon$ we find

\begin{equation}
H_2 ({\cal R})=m\pi
\left\{
{{\sin\left( \nu\log{({\cal R})}+\phi_2 -\arctan {1\over{2\nu}} \right)}\over
{\sin\left( \nu\log{({\cal R})}+\phi_2 +\arctan {1\over{2\nu}} \right)}}
\right\}.
\label{marginalmatchresult}
\end{equation}
Close to the zeroes of the right-hand side of this equation we can write instead
$\sqrt{\lambda_S }{\cal R} 
=(m+1/2)\pi +\epsilon$ and following the same procedure we find, to 
leading order in $\epsilon$,

\begin{equation}
H_2 ({\cal R})=(m+1/2)\pi - {1\over(m+1/2)\pi}  \left\{ {1\over 2} - 
{\nu}\tan\left( {{\nu}}\log{({\cal R})}+\phi_2\right)\right\}.
\label{marginalmatchresult2}
\end{equation}
A numerical solution of Eq.~(\ref{marginallogder}) shows that the approximate
solutions Eqs.~(\ref{marginalmatchresult},\ref{marginalmatchresult2}) are very
good within their ranges of validity. The expression in
Eq.~(\ref{marginalmatchresult2}) interpolates between two successive branches
of Eq.~(\ref{marginalmatchresult}) (see Fig.~\ref{running2}).

The three-body problem with short-range interactions is known to be equivalent,
in the ultraviolet regime, to the $1/r^2$ potential~\cite{efimov}.  The role of
the interparticle distance $r$ is played in the three-body problem by a
collective coordinate that vanishes when the three particles occupy the same
point in space, and the analogue of the three-body force is the short-distance
potential.  Since renormalization depends only on the short-distance behavior
of the theory, it is not surprising that the renormalization of the three-body
problem requires the presence of a three-body force~\cite{bosons}.
Using momentum-cutoff regularization the running of the three-body force was
found to follow Eq.~(\ref{marginalmatchresult}) very closely, even where this
formula predicts a pole. Evidently, for values of ${\cal R}$ where
Eq.~(\ref{marginalmatchresult2}) takes over in the $1/r^2$ problem, the
three-body force continues to follow Eq.~(\ref{marginalmatchresult}) and seems
to reach arbitrarily high values. Also, no evidence of multiple branches was
found in the three-body problem.  These discrepancies between the three-body
and the $1/r^2$ problem may be due to the fact that not all aspects of the
renomalization-group flow are universal, in the sense of being independent of
the particular regulator used.

\subsection{The Full Solution}

In the case $n=2$, the Schr\"odinger equation can be solved exactly for
all energies. The solution is 

\begin{equation}
u(x;\eta)= \sqrt{x}\left[ \exp{(i\alpha)} J_{i\nu}(\eta x) + 
\exp{(-i\alpha)} J_{-i\nu}(\eta x)\right]
\label{margx}
\end{equation}
where the $J_{\pm i\nu}$ are Bessel functions, and $\alpha$ is 
to be fixed by a boundary condition. For small $x$ we find

\begin{equation}
u(x;\eta)= \sqrt{x}\cos{\left( \nu\log{(x\eta/2 )}+\alpha-{\rm Im}\log
  \Gamma(1+i \nu)\right)}.
\label{asymtwcorr}
\end{equation}
Matching to Eq.~(\ref{marginalsmx}) gives

\begin{equation}
\phi_2=\alpha +\nu\log\eta/2-{\rm Im}\log
  \Gamma(1+i \nu) .
\label{marginalphieq}
\end{equation}
Since $\phi_2$ is, by construction, energy independent, $\alpha$ is
energy dependent.  

We can now look for solutions with $\eta=i\kappa$
which fall off exponentially at large $x$. It follows from
Eq.~(\ref{margx}) that

\begin{equation}
u(x;\kappa)\rightarrow \frac{1}{2}\exp{(i{\pi\over 4})}
\cos(\alpha+i\frac{\nu\pi}{2}) 
\exp{(\kappa x)} 
+ C\exp{(-\kappa x)}
\label{marglbssol}
\end{equation}
where $C$ is an energy-dependent coefficient.  
The bound-state solutions then correspond to 
$\alpha (\eta) =(m+1/2)\pi-i\nu\pi/2$, with $m$ an integer. 
Comparing with Eq.~(\ref{marginalphieq}) gives
the bound-state spectrum

\begin{equation}
E_m =-{2\over{M{r_0^2}}}\exp{\left(2
{{{\phi_2+{\rm Im}\log\Gamma(1+i\nu)-(m+1/2)\pi}}\over\nu}
\right)}.
\label{margbsspec1}
\end{equation}
Once $\phi_2$ is fixed by a single bound-state energy, all other energies
are predicted~\cite{case}. Adjacent bound-state energies are related by

\begin{equation}
{{\kappa_{m+1}}\over{\kappa_{m}}}=\exp{(-{\pi\over\nu})}.
\label{margbsspec2}
\end{equation}
Hence we see that the periodicity in the running coupling $H_2 ({\cal
R})$ is associated with the accumulation or dissipation of bound states
near the origin.

One can also fix $\phi_2$ to a scattering observable, like the scattering
length or the phase shift at a given energy.  Unfortunately, as for the Coulomb
potential, the $n=2,3$ singular potentials suffer infrared problems at low
energies, and therefore scattering lengths can be defined only if an infrared
cutoff is imposed~\cite{singpotreview}.

\section{Pure Singular Potentials: $n\geq 3$}

\subsection{The $k=0$ Solution}

The exact zero-energy solution for $n\geq 3$ is well
known~\cite{singpotreview}. Defining $z=\sqrt{\lambda_L}x^{1-n/2}/|1-n/2|$ and
$\phi (z)=u(x;0)/\sqrt{x}$, for $x > {\cal R}$ Eq. (\ref{sch1}) becomes an
ordinary Bessel equation:

\begin{equation}
\phi''(z) + \frac{1}{z}\phi' (z)
              + \left(1-\frac{1}{(n-2)^2 z^2}\right)\phi (z)=0.
\end{equation}
The solution is 

\begin{equation}
u(x;0)= \sqrt{x}\left[ A_n J_{1/(n-2)}\left(\frac{\sqrt{\lambda_L}}{1-n/2}x^{1-n/2}\right)
         +B_n J_{-1/(n-2)}\left(\frac{\sqrt{\lambda_L}}{1-n/2}x^{1-n/2}\right) \right],
\label{purezeroexact}
\end{equation}
which is a linear combination of Bessel functions. For small $x$ we can
write\footnote{In the case $n=4$, the Bessel functions are of half-integral
order and Eq.~(\ref{purezeroapp}) is exact for all $x$.}

\begin{equation}
u(x;0)={x^{{n/4}}}\cos\left({\sqrt{\lambda_L}\over{{n/2}-1}}
 {x}^{1-{n/2}} +\phi_n\right)\;\left[ 1\; +\; {\cal O}(x^{n/2-1})\right],
\label{purezeroapp}
\end{equation}
where we have set the constant prefactor to unity and

\begin{equation}
\phi_n = -{{n\pi}\over 4}{1\over{(n-2})} + 
i\log{\left(1+{{B_n}\over{A_n}}\exp{\left(-{{i\pi}\over{(n-2)}}\right)}\right)}.
\label{nphasedefined}
\end{equation}
This solution exhibits precisely
the same pathologies as Eq.~(\ref{marginalsmx}).

\subsection{Matching to the Square Well}

\begin{figure}[t]
\centerline{{\epsfxsize=4in \epsfbox{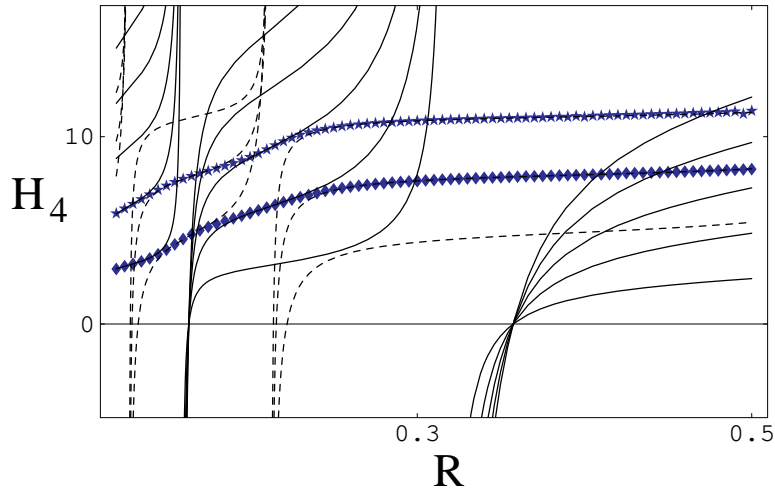}} }
\noindent
\caption{\it 
The running coupling for the $n=4$ singular potential. 
The solid line is given by Eq.~(\ref{matchresult}) and
the dashed line is given by Eq.~(\ref{matchresult2}). 
The triangles and stars are extracted from numerically
solving the Schr\"odinger equation and coincide exactly
with the numerical solution of Eq.~(\ref{logderivative}),}
\label{pretty2}
\vskip .2in
\end{figure}

We proceed as in the case $n=2$. Again we have a square well in the interior
region. Matching logarithmic derivatives at the boundary $x={\cal R}$ gives

\begin{equation}
\sqrt{\lambda_S}\cot{\sqrt{\lambda_S }{\cal R}}={n\over{4{\cal R}}} - 
\left({{\lambda_L}\over{{\cal R}^n}}\right)^{1/2}\tan \left({\sqrt{\lambda_L}\over{{n/2}-1}}
 {\cal R}^{1-{n/2}} +
\phi_n\right)
\label{logderivative}
\end{equation}
where we have neglected ${\cal O}({\cal R}^{n/2-1})$ corrections to the
wavefunction at $x>\R$.  The phase $\phi_n$ is physical and can be traded for
the scattering length (for $n>3$), as will be seen below.  If we vary ${\cal
  R}$ and $\lambda_S({\cal R})$ as given here, the phase $\phi_n$ will not be
affected.  We proceed as we did before in the $n=2$ case and find, in the
regions where the right-hand side of Eq.~(\ref{logderivative}) is large

\begin{equation}
H_n ({\cal R}) = {m\pi}
         \left\{1- \frac{1}
{1-n/4+\sqrt{\lambda_L}{\cal R}^{1-n/2}\tan\left(\frac{2\sqrt{\lambda_L}}{n-2}{\cal R}^{1-n/2}+\phi_n\right)}
\right\}.
\label{matchresult}
\end{equation}
In the other regime, where the right-hand side is close to a zero,
we have
\begin{equation}
H_n ({\cal R}) ={(m+{1\over 2})\pi}-{1\over {(m+{1\over 2})\pi}}
\left ( n/4-\sqrt{\lambda_L}{\cal R}^{1-n/2}\tan\left(\frac{2\sqrt{\lambda_L}}{n-2}{\cal R}^{1-n/2}+\phi_n\right)   \right).
\label{matchresult2}
\end{equation}
A numerical solution of Eq.~(\ref{logderivative}) in the case $n=4$ shows that
the approximate solutions Eqs.~(\ref{matchresult},\ref{matchresult2}) are very
good within their ranges of validity. The expression in
Eq.~(\ref{matchresult2}) interpolates between two successive branches of
Eq.~(\ref{matchresult}) (see Fig.~\ref{pretty2}).

The scattering length can be found from the zero-energy wavefunction for $n\geq
4$~\cite{perelomov}. For instance, we find the $n=4$ scattering length

\begin{equation}
a_4={r_0}\sqrt{\lambda_L}\tan\phi_4 .
\label{scattlnfour}
\end{equation}
It is evident that $a_4$ determines the phase $\phi_4$.

\section{The WKB Approximation} 

There is an important shortcoming in what we have done so far.  Defining
zero-energy scattering is {\it not} sufficient to guarantee correct
renormalization.  We want physics at all energies $\eta \simle 1$ to be cutoff
independent.  It is clear that the above procedure could in principle be
repeated at each energy by allowing energy dependence in $\lambda_S$. Fixing
$\phi_n$ at one energy and then predicting other energies will only result if
the scale of this energy dependence is much slower than $1/2Mr_0^2$, so that,
to some accuracy, $\lambda_S$ can be taken to be energy {\it in}dependent.
Otherwise, an infinite number of parameters (the strength of
arbitrarily-many-derivative contact interactions) would have to be known in
order to have predictive power.  This shortcoming can be removed using the WKB
approximation.  We can also consider the more general case, $f(x)\neq 1$.

\subsection{The WKB Criteria}

We now keep $f(x)$ arbitrary and consider the region $x>{\cal R}$ in the limit
${x}\rightarrow 0$.  A particularly well-suited approximation in this limit is
the WKB approximation, which is valid when the wavelength $\lambda$ is small
compared to the characteristic distance over which the potential varies
appreciably. That is

\begin{equation}
{1\over{2\pi}}\bigl|{{d\lambda}\over{dr}}\bigr|=|{{d}\over{dr}}[2M(E-V(r))]^{-1/2}|\ll 1 ,
\end{equation}
which translates into the constraint

\begin{equation}
{2\over n} \sqrt{\lambda_L f(x)}\gg x^{n/2-1}
\end{equation}
in the small-$x$ region\footnote{The WKB approximation is also valid
at large $x$ and finite $\eta$ provided that 
$\eta^3\gg n \lambda_L x^{-n-1}/2$.}.
Clearly this condition is satisfied for all $n>2$ as $x\rightarrow
0$. In the marginal case $n=2$ this condition is satisfied only for a
sufficiently strong potential. Therefore the WKB criterion parallels
the general argument given above based on the boundedness of the
Hamiltonian. The general WKB solution~\cite{landau} is

\begin{equation}
u(x;\eta )= \left( \eta^2 + {{{\lambda_L}{f(x)}}\over{x^n}}\right)^{-1/4}
\cos\left(\int_{x_0}^x dx' \left( \eta^2 + {{{\lambda_L}{f(x')}}\over{x'^n}}\right)^{1/2}\right)    
\label{genwkb}
\end{equation}
where $x_0$ is a constant of integration. For $V\gg E$ this reduces to

\begin{equation}
u(x;0)=  x^{n/4} f^{-1/4}(x)
\cos\left(\sqrt{\lambda_L}\int_{x_0}^x dx' x'^{-n/2} f^{1/2}(x')\right).    
\end{equation}
In the limit ${\cal R}<x\ll 1$, we can set $f(x)=1$ (keep the leading term in a
power series in $x$).  We then recover, for $n>2$,

\begin{equation}
u(x;0)= x^{n/4}
\cos\left( \frac{\sqrt{\lambda_L}}{n/2-1} x^{1-n/2} + \phi_n \right) 
\label{zerothWKB}   
\end{equation}
where $\phi_n ={-{\sqrt{\lambda_L}{x_0}^{-n/2+1}}/{(1-n/2)}}$.  The case $n=2$
is also recovered if one takes $\lambda_L\rightarrow \lambda_L -1/4$.
Therefore, we expect our conclusions about the renormalization of the singular
potentials to be valid for the more general case $f(x)\neq 1$.

\subsection{The Leading Energy Dependence}

We now show that the zero-energy solution is in fact sufficient to remove
cutoff dependence at all other low energies. The crucial point is that, in the
intermediate region ${\cal R}<x\ll 1$, for the energies of interest, the
potential energy is much larger than the total energy, and we recover the
zero-energy case.  This can be made more precise using WKB again \cite{case}.
We write the wavefunction for any $x\ge {\cal R}$ as

\begin{equation}
u(x;\eta)= A(x;\eta) u(x;0).
\label{edepwf}
\end{equation}
Then $A(x;\eta)$ obeys

\begin{equation}
\frac{d^2 A(x)}{dx^2}
+ 2 \frac{d \ln u(x;0)}{dx} \frac{d A(x)}{dx}
+\eta^2 A(x)=0 
\end{equation}
which depends only on the zero energy wavefunction.
Now, since for ${\cal R}<x\ll 1$,

\begin{equation}
\left|\frac{d \ln u(x;0)}{dx}\right| \gg 1,
\end{equation}
$A(x)$ can be written

\begin{equation}
A(x) = A_{(0)}(x) +A_{(1)}(x) +\ldots,
\end{equation}
where 

\begin{equation}
\frac{dA_{(0)}}{dx} =0;\qquad
\frac{dA_{(1)}}{dx} =-{1\over 2}\frac{u(x;0)}{u'(x;0)} \eta^2 A_{(0)};
\qquad \ldots
\end{equation}
We then find the leading energy corrections

\begin{eqnarray}
A(x;\eta) = A_{(0)} 
\left\{ 1-\frac{\eta^2}{2} 
          \int_0^x dx'\frac{u(x';0)}{u'(x';0)} +\ldots \right\}
\label{kwkb}
\end{eqnarray}
We see that, in the intermediate region, the energy dependence of the
wavefunction (\ref{edepwf}) is determined by the zero-energy wavefunction
$u(x;0)$.  If the phase of $u(x;0)$ has been fixed, the phase of $u(x;\eta)$ is
fixed, and scattering observables can be predicted at low energies.

\subsection{Error Estimates}

The fact remains that our arguments are all at short distances where the WKB
approximation is valid. This is, of course, the opposite of the EFT limit which
interests us. One may wonder whether cutoff effects can be amplified when
propagating the wavefunction from short to long distances. We will see now that
this cannot occur and in turn find an estimate of the cutoff error associated
with the scattering phase shift. Usually, in perturbative EFT, the error is a
power law in $\R$. Here we will find a more complicated functional dependence.

By adjusting $H_n({\cal R})$ as in Eqs.~(\ref{matchresult},\ref{matchresult2})
we guarantee that two zero-energy solutions $u_\R(x;0)$, $u_\Rp(x;0)$
corresponding to two different cutoffs ${\R},\R'\ll 1\sim 1/\eta$ are
identical.  At finite values of $\eta$, solutions obtained with different
cutoffs will no longer be equal, but their difference can be easily estimated.
Taking ${\cal R'}<{\cal R}$, the Schr\"odinger equations satisfied by
$u_\R(x;\eta )$ and $u_\Rp(x;\eta )$ are the same in the $x>\R$ region so their
Wronskian,

\begin{equation}
W[u_\R,u_\Rp](x;\eta )=u_\R(x;\eta )  u_\Rp'(x;\eta ) - u_\R'(x;\eta )
u_\Rp(x;\eta )
\label{W} 
\end{equation}
is independent of $x$. At large distances ($r\gg
(\lambda_L/k^2)^{(1/n)}$), where the solutions are plane waves,
$W[u_\R,u_\Rp]$ is related to the phase shifts $\delta_\R,\delta_\Rp$
obtained with the cutoffs $\R$ and $\Rp$ by

\begin{equation}
W[u_\R,u_\Rp](r\gg (\lambda_L/k^2)^{(1/n)};\eta)=
A_\R A_\Rp \eta \sin(\delta_\R-\delta_\Rp), 
\label{Wlarge}
\end{equation}
where $A_\R, A_\Rp$ are the amplitudes at large distances. These prefactors
are easily estimated from the general WKB solution, Eq.~(\ref{genwkb}), in
the region $r\gg (\lambda_L/k^2)^{(1/n)}$, where the WKB solution maps
to the asymptotic plane-wave solution (see Footnote 9). We find
$A_\R\, , A_\Rp\sim \eta^{-1/2}$.

On the other hand, at the cutoff distance $x=\R$, $W[u_\R,u_\Rp]$ is estimated
using our WKB formula, Eq.~(\ref{kwkb}). We find

\begin{equation}
W[u_\R,u_\Rp]({\cal R};\eta )= 
W[u_\R,u_\Rp]({\cal R};0)-{{\eta^2}\over 2} {\cal E}({\cal R};0 ) +\ldots 
\end{equation}
where

\begin{equation}
W[u_\R,u_\Rp]({\cal R};0)=u_\R({\cal R};0) u_\Rp({\cal R};0)
\left[\frac{u_\Rp'({\cal R};0)}{u_\Rp({\cal R};0)}
-\frac{u_\R'({\cal R};0)}{u_\R({\cal R};0)}+
{\cal O}\left({\cal R}^{n/2-1}{\frac{u_\R'(\R )}{u_\R(\R )}}\right)\right] 
\label{zerowr}
\end{equation}
and 

\begin{eqnarray}
{\cal E}({\cal R};0 )&\equiv&
u_\R({\cal R};0) u_\Rp({\cal R};0)
\left[\frac{u_\Rp({\cal R};0)}{u_\Rp'({\cal R};0)}
-\frac{u_\R ({\cal R};0)}{u_\R'({\cal R};0)}
+{\cal O}\left({\cal R}^{n/2-1}{\frac{u_\R(\R )}{u_\R'(\R )}}\right)\right] \nonumber \\
&+&W[u_\R,u_\Rp]({\cal R};0)
{\int_0^{{\cal R}}}dx'
\left[\frac{u_\Rp(x';0)}{u_\Rp'(x';0)}
+\frac{u_\R (x';0)}{u_\R'(x';0)}
+{\cal O}\left({x'}^{n/2-1}{\frac{u_\R(x')}{u_\R'(x')}}\right)\right].
\end{eqnarray}
We have included the error due to keeping only the leading zero-energy
wavefunction in Eq.~(\ref{purezeroapp}).  Recall that we choose our fitting
procedure to be energy independent, for example, by comparing the zero-energy
wavefunction to the scattering length.  It then follows that
$W[u_\R,u_\Rp]({\cal R};0)=0$, by construction, for the full wavefunction, and
from Eq.~(\ref{zerowr}) we have

\begin{equation}
\frac{u_\Rp'({\cal R};0)}{u_\Rp({\cal R};0)}-\frac{u_\R'({\cal R};0)}{u_\R({\cal R};0)}=
{\cal O}\left({\cal R}^{n/2-1}{\frac{u_\R'(\R )}{u_\R(\R )}}\right).
\label{zerowrin}
\end{equation}
Using these constraints it is straightforward to find

\begin{equation}
{\cal E}({\cal R};0 )= {\cal O}\left({\cal R}^{n/2-1}
\frac{{(u_\R({\cal R}))^3}}{u_\R'({\cal R})}\right).
\label{zerowrinb}
\end{equation}
If we assume that all oscillating functions of $\R$ are of order unity for
values of $\R$ at which we fit observables, then ${\cal E}({\cal R};0 )= {\cal
  O}( {\cal R}^{3n/2-1})$, which is small for all $n\geq 2$.  Matching the
Wronskians at large ($r\gg (\lambda_L/k^2)^{(1/n)}$) and short ($x=\R$)
distances then yields an estimate for the error in the phase shift:

\begin{equation}
\delta_\R-\delta_\Rp \sim \eta^2 \ \ {\cal E}({\cal R};0 )
\label{pherror}
\end{equation}
where ${\cal E}({\cal R};0 )$ is a function of $\cal R$ whose complicated
parametric cutoff dependence is given by Eq.~(\ref{zerowrinb}).  This shows
that the renormalization procedure described here produces cutoff independent
phase shifts, accurate up to order $\eta^2 \ {\cal E}({\cal R};0 )$.

\section{The Weak Coupling Limit}

It might seem odd that the explicit dependence on the coupling constant is
nonanalytic in the formula for the $n=4$ scattering length,
Eq.~(\ref{scattlnfour}). Naively it would appear that nonperturbative effects
are important at arbitrarily weak coupling.

However, we know that this cannot be the case, since for weak coupling the
scattering length should go smoothly to its square-well value.  We would expect
a perturbative description in the singular potential to be valid when the
potential energy, $-{\lambda_L} {x^{-n}}$, is much smaller than the kinetic
energy, ${x^{-2}}$. This leads to the condition

\begin{equation}
r\gg {r_0}\lambda_L^{1\over{n-2}},\quad {\rm or}\qquad
k\ll {r_0^{-1}}\lambda_L^{-{1\over{n-2}}}.
\end{equation}

In effect, taking $\phi_4$ from Eq.~(\ref{logderivative}) 
we find, for $\lambda_L/ \R^2\ll 1$,

\begin{equation}
\frac{a_4}{R}=\left(1 - \frac{\tan ({\sqrt{{\lambda_S}}}\,{\cal R})}{{\sqrt{{\lambda_S}}{\cal R}}}\right) 
-\frac{1}{3}\left( 1+ \frac{\tan ({\sqrt{{\lambda_S}}}\,{\cal R})}{{\sqrt{{\lambda_S}}{\cal R}}}+\left(\frac{\tan ({\sqrt{{\lambda_S}}}\,{\cal R})}{{\sqrt{{\lambda_S}}{\cal R}}}\right)^2\right)
{{\lambda_L}\over{\R^2}} +
  {{\cal O}( {{{\lambda_L}^2}\over{\R^4}} )}.
\end{equation}
Leading order reproduces the square-well scattering length and the corrections
are analytic in $\lambda_L$. Hence, there is, in fact, no nonanalyticity near
zero coupling in the presence of the square well.

\begin{figure}[t]
\centerline{{\epsfxsize=3.6in \epsfbox{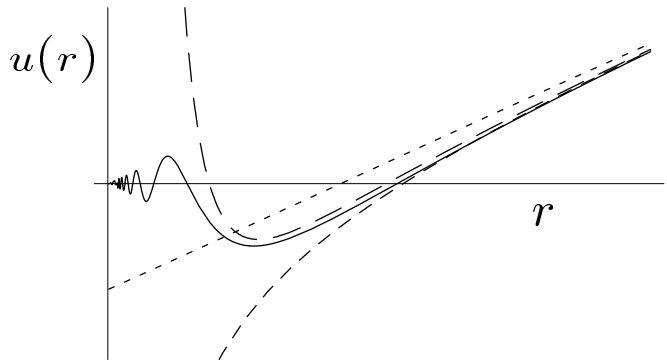}} }
\noindent
\caption{\it The exact zero-energy wavefunction (solid line)
at small $r$ compared 
with the wavefunction
obtained in perturbation theory to
leading order 
(small dashes), next-to-leading order 
(medium dashes), and  next-to-next-to-leading order 
(large dashes).}
\label{perturb}
\vskip .2in
\end{figure}

Of course, if the cutoff $\cal R$ is taken at values where the oscillatory
behavior of the wavefunction has set in, $\R \simle \lambda_L^{1\over{n-2}}$,
then there is no sense in which perturbation theory in $\lambda_L$ can capture
the true behavior of the wavefunction. This is made clear in Fig.~
\ref{perturb} where several orders in a perturbative expansion of the $n=4$
singular-potential wavefunction are plotted against the exact
singular-potential wavefunction in the short-distance region.

\section{Numerics}

In this section we analyze the $1/r^4$ potential numerically.  For simplicity,
we take $2M={r_0}=1$; therefore, $x=r$ and $\eta =k=\sqrt{2E}$. The
long-distance potential is then completely determined by $\lambda_L$ which we
take to be unity.  We consider the ``natural case'', which is characterized by
$a_4\sim (\lambda_L )^{1/2}$, and the ``unnatural case'', which is
characterized by $a_4 \gg (\lambda_L )^{1/2}$.

\begin{figure}[t]
\centerline{{\epsfxsize=3.4in \epsfbox{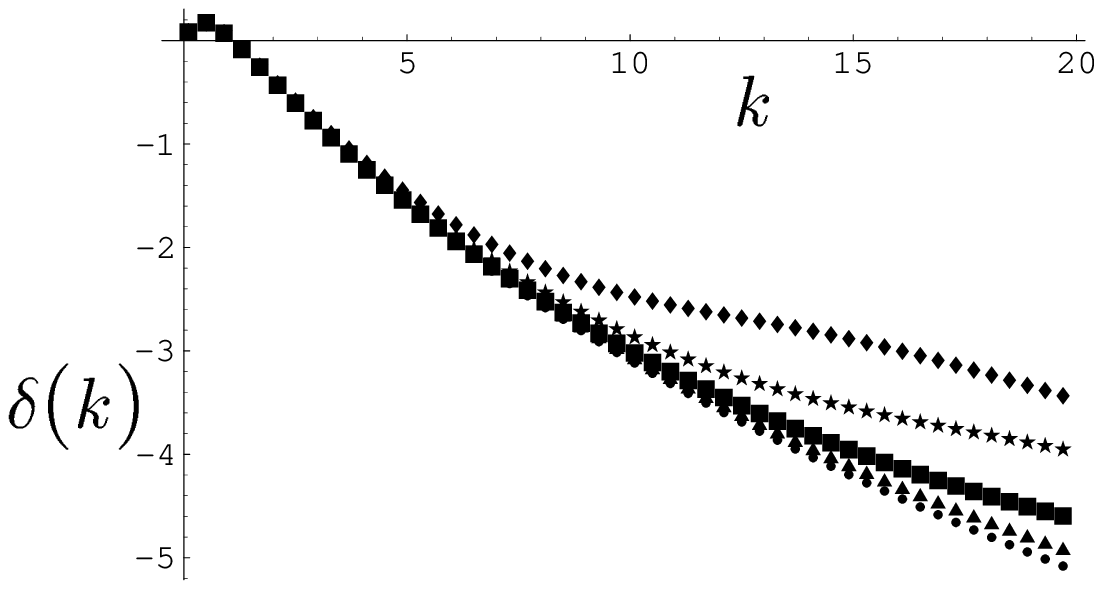}} {\epsfxsize=2.9in \epsfbox{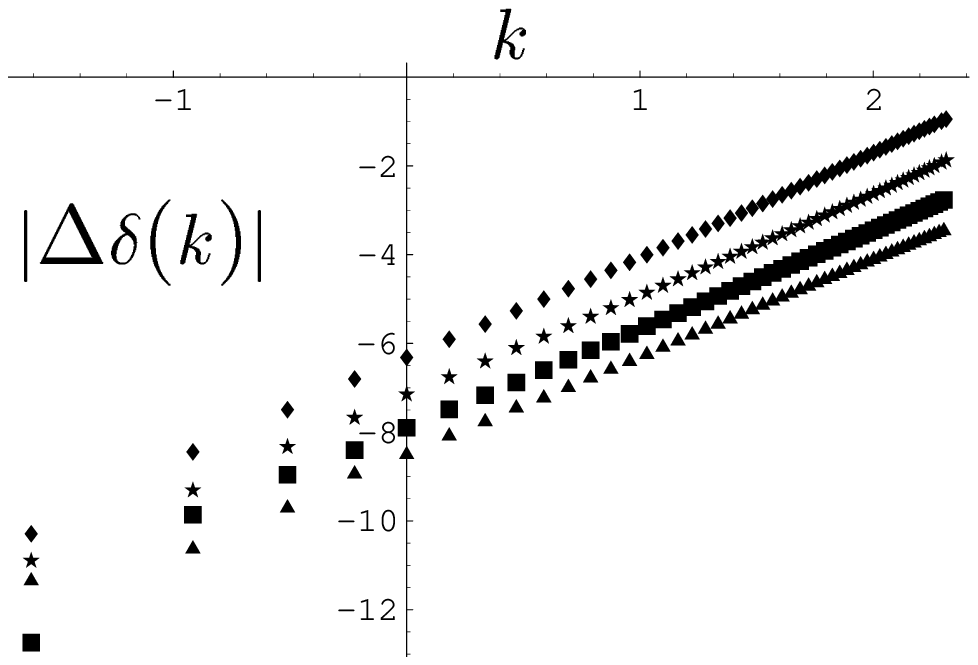}}} 
\noindent
\caption{\it 
  Left: phase shifts $\delta$ vs. energy in the case of a natural scattering
  length for $n=4$. Various cutoffs are given with the square well tuned to
  give the same scattering length. The cutoffs are $R=.01$ (dots), $R=.02$
  (triangles), $R=.04$ (squares), $R=.08$ (stars), $R=.16$ (diamonds).  Right:
  logarithm of the errors $|\Delta\delta (k)|=\delta_R-\delta_{R'}$ as a
  function of the logarithm of the energy for $n=4$.  The pairs of cutoffs
  $(R,R')$ are: $(.16,.08)$ (diamonds), $(.08,.04,)$ (stars), $(.04,.02,)$
  (squares), $(.02,.01,)$ (triangles).  }
\label{deltavsknat}
\vskip .2in
\end{figure}

\begin{figure}[t]
\centerline{{\epsfxsize=3.4in \epsfbox{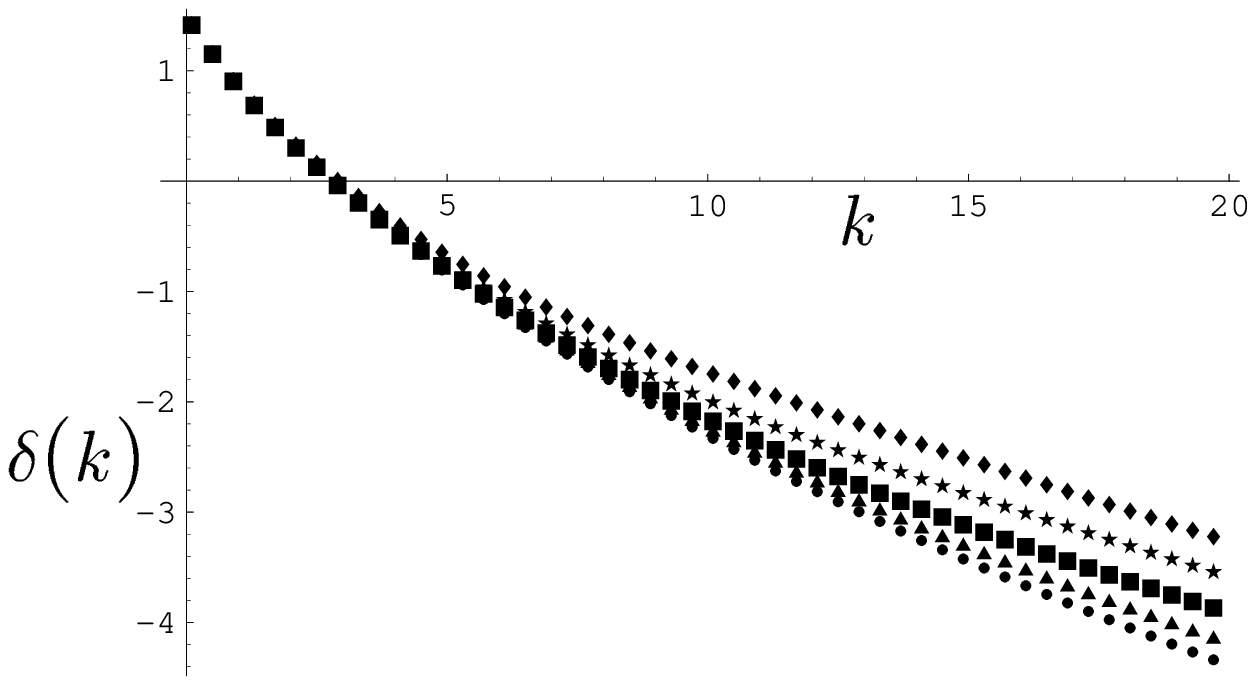}} {\epsfxsize=2.9in \epsfbox{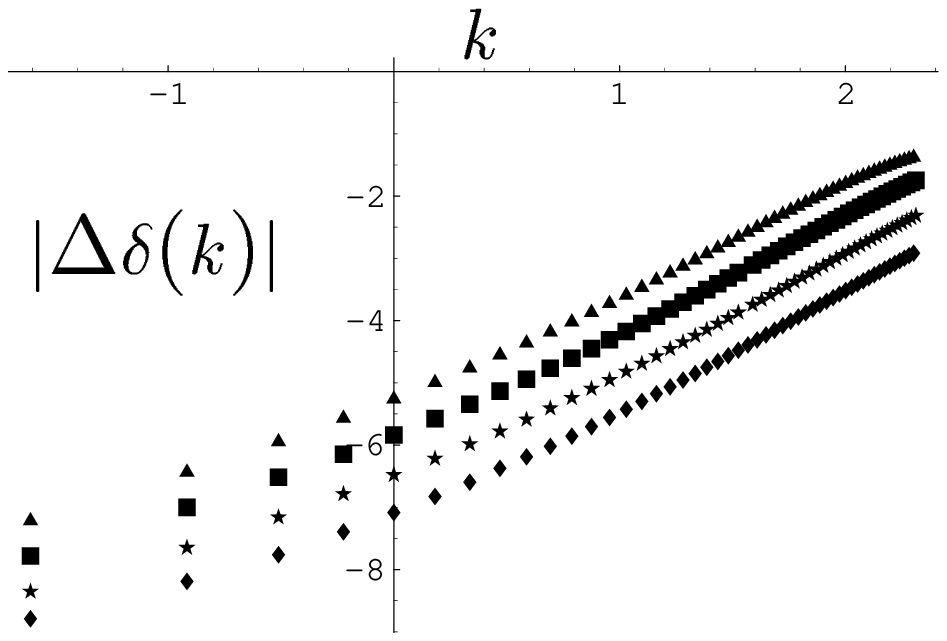}}}
\noindent
\caption{\it Same as Fig.~\protect\ref{deltavsknat} for an unnatural  
scattering length.}
\label{deltavskunnat}
\vskip .2in
\end{figure}

In Fig.~\ref{deltavsknat} we show phase shifts $\delta (k)$ in a natural case
($\delta (0.1)=0.1$ and $\phi_4 =-101.298$) for various cutoffs.  We see that,
as anticipated, the low-energy phases are to a good approximation cutoff
independent; cutoff dependence becomes more pronounced as the cutoff radius and
the energy are increased.  In the same figure we also plot the error analysis:
the (log of the) errors $|\Delta\delta (k)|=\delta_R -\delta_{R'}$ as a
function of (the log of the) energy for various pairs of cutoffs.  We find that
the errors scale as $k^2$, as expected on the basis of Eq.~(\ref{pherror}).  In
Fig.~\ref{deltavskunnat} we show the corresponding results in an ``unnatural''
case ($\delta (0.1)=\pi/3$ and $\phi_4 =-98.954$).  Again we find that the
errors scale as $k^2$.

\section{Conclusion}

We have reconsidered singular potentials of the form $1/r^n$ with $n\geq 2$
from the viewpoint of modern renormalization theory. We have shown that the
well-known pathologies near the origin are cured by a square-well counterterm
which represents the effect of unknown short-distance physics. The
renormalization-group evolution of this counterterm has periodic
behavior. The counterterm is not determined uniquely at any given cutoff due to
the infinite number of branches of the renormalization-group flow, and one is
allowed to freely jump from branch to branch at will without causing any change
in the low-energy phase shifts (up to ${\cal O}(k^2 )$ corrections).

The dependence of the number of bound states on the choice of branch is
complex. Arguments similar to the one leading to Eq.~(\ref{margbsspec1}) are
valid for a generic singular potential as long as (i) the binding energy is
much smaller that $\lambda_S/(2Mr_0^2)$ (in order that the binding energy can
be disregarded in the left-hand side of Eq.~(\ref{logderivative})) and (ii) the
binding energy is much larger than $\lambda_L/(2Mr_0^2)$ (so the fact that
$f(r/r_0)\neq 1$ is unconsequential).  The resulting spectrum shows a power-law
distribution that is given by the WKB estimate~\cite{perelomov}. 
We see now that as ${\cal R}\rightarrow
0$ and $\lambda_S({\cal R})\rightarrow\infty $ the region of validity of the
calculation sketched above grows and more and more bound states are created.
If, in addition to keeping the phase shifts cutoff independent, one also
demands that the number of bound states also be fixed, the value of the
counterterm $\lambda_S$ jumps down a branch at every cycle.  We are then left
with a periodic, limit-cycle behavior for the renormalization-group flow. This
is a unique situation in field theory/critical phenomena, where the standard
behavior corresponds to counterterms approaching either zero (asymptotic
freedom) or infinity as the momentum cutoff goes to infinity ~\cite{wilson}.

Naively it appears that the singular potentials have a nonanalytic dependence
on the coupling parameter even at weak coupling. This would negate a
perturbative description at weak coupling, a conclusion which must be
incorrect. One might imagine some as yet experimentally invisible light
particle which interacts at long distances via a singular potential (e.g. an
axion). If the behavior of the wavefunction were such that there is a branch
point at the origin of the coupling-constant plane, then nonperturbative
effects would persist even for couplings of gravitational strength. We have
seen that this nonanalyticity is an artifact which is removed by short-distance
physics encoded by the square well.

Renormalization renders low-energy phase shifts cutoff independent up to ${\cal
  O}(k^2 )$ corrections. The cutoff dependence of these errors is not generally
a power law as one expects in Wilsonian EFT. The renormalization-group flow
introduces complicated oscillatory behavior in the corrections, which
nonetheless is small for judiciously chosen cutoffs. Our theoretical
expectations of the error have been confirmed numerically.  We expect that the
methods developed in this paper will prove useful to those interested in the
cornucopia of physical systems whose long-distance behavior is governed by
singular potentials.

\vspace{1cm}
\noindent
{\large\bf Acknowledgments}

\noindent
We thank David Kaplan, Hans Hammer and Martin Savage for valuable conversations
and Sid Coon for bringing Ref.~\cite{perelomov} to our attention.  UvK thanks
the Nuclear Theory Group and the Institute for Nuclear Theory at the University
of Washington for hospitality, and RIKEN, Brookhaven National Laboratory and
the U.S. Department of Energy [DE-AC02-98CH10886] for providing the facilities
essential for the completion of this work.  This research was supported in part
by the DOE grants DE-FG03-97ER41014 (SRB) and DOE-ER-40561 (PFB), and by NSF
grant PHY 94-20470 (UvK).  LC and JMc are grateful to the University of
Washington REU program of the NSF for support.

\end{document}